\documentstyle[12pt]{article}
\newcommand{\be}{\begin{equation}}
\newcommand{\ee}{\end{equation}}
\newcommand{\ba}{\begin{eqnarray}}
\newcommand{\ea}{\end{eqnarray}}
\newcommand{\baa}{\begin{eqnarray*}}
\newcommand{\eaa}{\end{eqnarray*}}
\newcommand{\bb}{\begin{thebibliography}}
\newcommand{\eb}{\end{thebibliography}}
\newcommand{\Ds}{\displaystyle}
\newcommand{\ci}[1]{\cite{#1}}
\newcommand{\bi}[1]{\bibitem{#1}}
\begin {document}

\begin{center}
{\Large \bf The elastic contribution to the Burkhardt -- Cottingham and Generalized
Gerasimov--Drell--Hearn Sum Rules}\\[1cm]
{I.V. Musatov}\\[0.3cm]
{\it  Old Dominion University, Norfolk, USA  }
\footnote{On leave from: \it Institute of Applied Physics,
Tashkent State University, Tashkent, Uzbekistan 700095. }\\[0.3cm]
{O.V. Teryaev}\\[0.3cm]
{\it Bogoliubov Laboratory of Theoretical Physics\\
Joint Institute for Nuclear Research, Dubna\\
Head Post Office, P.O. Box 79, 101000 Moscow, Russia}\\[0.3cm]
and\\[0.3 cm]
{A. Sch\"afer}\\[0.3cm]
{\it Institut f\"ur Theoretische Physik\\
Universit\"at Regensburg\\
D-93040 Regensburg}
\end{center}
\begin{abstract}
The elastic contribution to the first moment of $g_2(x,Q^2)$ is
analysed using a Drell-Yan-West type of relation and is shown to be
negative.
For a qualitative estimate the 
one-loop contributions to the polarized DIS sum rules in QED are studied.
The behaviour of the generalized Gerasimov--Drell--Hearn
sum rule is sensitive to infrared regularization.
With a lower threshold for the gluon virtuality
the relation of the generalized GDH sum rule to the
Burkhardt--Cottingham sum rule is studied. 
We conclude that the elastic part has to be included
for long range interactions but can be consistently discarded
for short range interactions.  
\end{abstract}

\newpage

The generalized Gerasimov-Drell-Hearn (GDH) sum rule \ci{Ger,DH} is
just being tested experimentally \cite{E143} and
the available proton data are in good agreement with the predictions
made in
\cite{SoTe,ST95}, which used the relationship between GDH and Burkhardt-Cottingham (BC)
sum rule. They also agree with a new estimate of the
contributions from low-lying resonances \cite{Ioffe}.
We stress, that such a similarity is by no means surprising,
since the dominant magnetic form factor of $\Delta(1232)$,
being the main source of the rapid variation of GDH \cite{IoLi} is
contributing entirely through the structure function $g_2$ \cite{ST95},
which is the key ingredient to the approach of \cite{SoTe}.

The starting point of this approach  is the simultaneous analysis
of 
GDH and BC sum rules, inspired by the paper of
Schwinger \ci{Sch}. To verify the latter, the  check of BC and GDH
sum rules in QED was performed almost 20 years ago \ci{Mil} (although 
the BC sum rule
was not mentioned in this paper). In the present article we complete their
calculation and use the model QED case in order to make (qualitative)
statements about the behaviour of the GDH sum rule for the proton.
Also we try to clarify the role of the elastic contribution for 
$x\rightarrow 1$.

The main problem with the (generalized) GDH
sum rule is the following.
Let us introduce the $Q$-dependent integral \cite{AIL}
\ba
I_1(Q^2)&={2 M^2\over {Q^2}} \int^1_0 g_1(x,Q^2) dx
~=~ \int_{Q^2/2M}^{\infty} {d\nu\over \nu}~ G_1(\nu,Q^2) ~~,
\nonumber \\
I_2(Q^2)&={2 M^2\over {Q^2}} \int^1_0 g_2(x,Q^2) dx
=M^2 ~ \int_{Q^2/2M}^{\infty} d\nu~ G_2(\nu,Q^2)~~.
\label{I1}
\ea

defined for all $Q$.
There are solid theoretical arguments to expect a strong $Q^2$-dependence
of $I_2$. It is the well-known Burkhardt-Cottingham sum rule \ci{BC},
derived independently by Schwinger \ci{Sch}
with a rather different method. It states that
\be
I_2(Q^2)={1\over 4}\mu G_M (Q^2)[\mu G_M (Q^2) - G_E (Q^2)],
\ee
where $\mu$ is the nucleon magnetic moment and the  $G$'s are the familiar
Sachs form factors which are dimensionless and normalised to unity
at $Q^2=0$. For large $Q^2$ one can neglect the r.h.s. and gets
\be
\int^1_0 g_2(x) dx=0.
\ee

The latter equation is often called the BC sum rule
and applies only up to corrections of twist higher than four.

One of the crucial points of the whole discussion is the 
treatment of the elastic contribution. Being of high twist it is not
explicitely treated in standard OPE analyses. For the small $Q^2$
values relevant here we follow the arguments of \cite{SoTe,ST95}
which show that for kinematic reasons  if one requires a smooth
interpolation to $Q^2=0$.  
the elastic contribution at $x=1$
should not be included in the sum rule (\ref{I1}). One recovers then at $Q^2=0$
the GDH sum rule:
\be
I_1(0)=-{\mu_A^2 \over 4}.
\ee
where $\mu_A$ is the nucleon anomalous magnetic moment in nuclear magnetons.
While $I_1(0)$ is always negative, its value at large $Q^2$ is determined
by the  integral $\int^1_0 g_1(x,Q^2) dx$ and is thus
positive for the proton. This illustrates the existence of strong 
scaling violations for
$I_1$ in the
region $0 < Q^2 < 1~{\rm GeV}^2$.
Its origin can be elucidated somewhat using a modified Drell-Yan-West
relation \cite{DYW}. As for the unpolarized case one can relate the
elastic and quasielastic part of the first moment of the
spin-dependent
structure functions  to formfactors 
(at large $Q^2$) according
to
\be
\int_{1-cons/Q^2}^1 ~g_2(x,Q^2)dx 
= -{Q^2\over 2}~F_2(Q^2) \left( F_1(Q^2)+{F_2(Q^2)\over 2M} \right)
\ee
where the right hand side is just the elastic contribution. 
In the language of OPE it corresponds to 
contributions from cat-ear diagrams.
The constant $cons$ is left free. In QCD it would be proportional to
the duality interval ($\sim \nu M$). 
If one makes the usual ansatz for the form of
$g_2(x)\sim (1-x)^n$ for $x\rightarrow 1$ and uses the fact that 
$F_2(Q^2)\sim (1/Q^2)^3$
and $F_1(Q^2)\sim (1/Q^2)^2$ equation (3) gives $n=3$. This
prediction could be tested by planned SLAC and CEBAF experiments. It
is
of the same nature as the usual quark-counting rule predictions.\\

The role
of the elastic contribution is in principle quite similar 
for polarized and unpolarized structure functions the only difference is
that because the leading  contributions are zero the cat-ear contribution
is dominant for the first moment of $g_2(x)$. It leads to the
non-trivial prediction of the negative sign.\\

It is possible to decompose $I_1$ into the contributions from the two form factors
$I_{1+2}$ and $I_2$:
\be
I_1=I_{1+2}-I_2,
\ee
where
\be
I_{1+2}(Q^2)={2 M^2\over {Q^2}} \int^1_0 g_{1+2}(x,Q^2) dx
\ee
\be
g_{1+2}=g_1 + g_2.
\ee
Note that this decomposition corresponds to extracting coefficient functions
in front of two independent tensor combinations in the spin--dependent
(antisymmetric) part of the hadron tensor:
\be
W_{\mu\nu}^a \sim  g_{1+2} \, \epsilon_{\mu \nu \rho \sigma} q^\rho s^\sigma
                -  g_2 \,     \epsilon_{\mu \nu \rho \sigma} q^\rho p^\sigma .
\ee

For $Q^2\rightarrow0$ one finds
\be
I_2(0)={\mu_A^2+\mu_A e \over 4},
\ee
$e$ being the nucleon charge in elementary units.
To reproduce the GDH value one should have
\be
I_{1+2}(0)={\mu_A e \over 4}.
\ee
Note that $I_{1+2}$ does not differ from $I_1$ for large $Q^2$ because the
BC sum rule holds there and that it is {\it positive} for the proton. 
A smooth interpolation for  $I_{1+2}(Q^2)$
between large $Q^2$ and $Q^2=0$ can be found in \ci{SoTe}.

To better understand the issue of 
the GDH sum rule
it seems reasonable to investigate its generalized version
in a simple theory,
such as a perturbative gauge field model.

An implication for the BC sum rule comes from the check of sum rules
(7) and  (10) in QED performed immediately after Schwinger's paper \ci{Mil}.
This pioneering paper is hardly known in the spin community, probably,
due to two main reasons.
First, it uses the Schwinger sources theory,
which is actually unproblematic in this case, because the result are the same
in the diagrammatic approach, as we shall show below.
Second, they use  unconventional 
definitions for spin structure functions (also first introduced by
Schwinger, e.g. $H_4$ stands for $g_2$).

Note that it is better to speak here about "perturbative QCD", just
because the emission
of a real photon by the Bethe-Heitler process  (not taken into account in
\ci{Mil}) is of the same order as its emission by the "internal" quark.
To exclude the former, one may change this photon to a gluon. Since we consider
only the first order in $\alpha$, the result is the same apart from a trivial
color factor.
The BC sum rule in such an approximation is obviously valid, if the elastic
contribution is included into the integrand \ci{Mil}. 

To proceed, we note that in a gauge theory (for definiteness,
we will refer to perturbative QCD), while both parts
(elastic and inelastic) of the BC sum rule are infrared stable,
the generalized version of the GDH sum rule is infrared
divergent (logarithmically) at any nonzero $Q^2$.
It is well--known, however, that the infrared divergencies
coming from the elastic
process cancel those in the inelastic one, implying
that the quantity
\be
\tilde I_1(Q^2) = I_1(Q^2) + I_1^{elastic}(Q^2)
\ee
is infrared finite. Unfortunately, it diverges as $1/Q^2$ at small $Q^2$ due
to kinematic factors in $I_1^{elastic}(Q^2)$.
At the same time, the Born contribution to $I_2$ is zero. The IR
divergent part of the elastic
contribution has the kinematic structure of the Born term, so it is also zero in the
case of $I_2$ making the inelastic contribution to be IR finite.

The physical reason for this problem is that the possibility of
emiting soft gluons
contradicts the existence of a finite threshold, which is
assumed in the original version of the GDH sum rule.
So, the generalized GDH sum rule needs to be defined more carefully 
for gauge models. Possible solutions could be a finite mass of the gluon
or suppression of  soft gluons with virtualities less than some value
$\lambda^2$, which can be interpreted as the threshold of 
detector sensitivity.

In this paper we will use the latter option for the regularization of
IR singularities
in $I_{1+2}$. The calculation with regularization by a finite gluon mass is
more complicated, and the result clearly should be the same.

Explicit calculations gives the following expression:
\be
I_i=\int_0^{1-\delta} dx g_i(x), \ \ \ \ \delta = {2 m \lambda \over Q^2}
\ee

\ba
g_{1+2} &=& {1 \over 2} \delta (1-x)    \nonumber  \\
        & & + {1\over 2} {\alpha_s \over 4 \pi} C_F
            \left \{
            - { 5 y^2 + 2 y^2 x + 6 y x - 11y^2 x^2 + 36 y x^2 + 4 y^2
x^3 - 34 y x^3 + 32 x^3
                                      \over
                           ( y + x - x y ) ( y + 4 x^2 ) ( 1 - x)   }
\right. \nonumber \\
         & &  \left.
            - 2 { - 2 y^2 x - 3 y x + y^2 x^2 - 10 y x^2 - y^2 + y x^3 -
            16 x^3
                                       \over
                    (1-x) (y + 4 x^2) \sqrt{ y (y + 4 x^2) } }
          \cdot \log {\cal D}
          \right \}, \\[3mm]
{\cal D} &=& {2x +y +\sqrt{y (y+4x^2)} \over  2x +y -\sqrt{y (y+4x^2)} };
     \ \ \ \ \ y={Q^2 \over m^2} .
\label{g12}
\ea

Below we reproduce also the full expression for $g_2$
obtained first in the paper \ci{Mil}
(see also the more recent calculations \cite{Alt}):
\be
\begin{array}{rl} \Ds
g_2= {\alpha_s \over 2 \pi} {1 \over 4} C_F {y \over 2}
      & \Ds \left \{ {2 \over 1+y} \left [ \Ds {y \over (y+x-xy)^2}
      - {6x(2(3y+2)x+2y+3)y) \over (y+4x^2))^2}
           \right. \right. \\
      & \Ds   \left.  \left. \ \ \ \
      - {1 \over y+4x} - {x(y^2+1) \over (y+x-xy) (y+4x^2)} \right ]
                   \right. \\
      & \Ds   \left.
      + \left ( 1+2y- 3x {4x(y-1)-5y \over y+4x^2} \right )
                                                  {4x^2\over \sqrt{y}
(y+x-xy)^{3/2}}
                                                  \log {\cal D} \right \},
\label{g2exact}
\end{array}
\ee
For QED  $m$ is the mass of the Dirac particle. It corresponds 
to some constituent quark mass in QCD.

 Comparison of this expression for $g_2$ with the asymptotic (in the
limit $Q^2 \to \infty$)
formula obtained in \ci{Mer-vN} immediately shows that the term omitted in that
paper is the first term in the expression above:
\be
\Delta g_2 = {\alpha_s \over 2 \pi} {1 \over 4} C_F  \cdot
           {1 \over y} \cdot {1\over \left [ (1-x)(1-1/y) + 1/y \right ]^2}
\ee
and naively suppressed at high $Q^2$ as $m^2/Q^2$. However, due to
the (integrated)
singularity in the denominator, it should be taken into account for
moments
and leads to
the correct result for the BC sum rule.
More precisely, this term is proportional to
\be
\lim_{\epsilon \to 0} \ \  {\epsilon \over (\bar x+\epsilon)^2} =
                               \delta_+ (\bar x)  ;
\ \ \ \ \ (\bar x = 1-x, \ \ \ \ \epsilon = 1/y),
\ee
which looks like the {\em elastic} contribution (cf Ref. \cite{Alt}).

It is interesting to investigate this term in the opposite limit of
very small $Q^2$:
\be
{y \over [x(1-y) + y]^2} \to {\epsilon \over (x+\epsilon)^2} \sim \delta_+(x);
\ \ \ \ (\epsilon = y).
\ee
It can be easily found that its contribution to the BC sum rule at
$Q^2=0$ is $2 I_2(0)$,
i.e. without it one would obtain the
correct absolute value but the wrong sign.

The numerical results for the generalized GDH sum rule in the lowest order
are shown for different values of
the IR cut-off parameter in Fig.1. It can be seen that smaller values of
$\delta$ lead to higher IR peaks
closer to the abscissa. In the opposite regime of large $\delta$
(which effectively is expected in QCD)
it shows a rather smooth behavior, which is compatible with what was predicted
for the
generalized sum rule $I_{1+2}(Q^2)$ for the proton in \cite{SoTe}

In conclusion, we presented here the investigation of the generalized
Gerasimov-Drell-Hearn sum rule in the framework of perturbation theory.
This simple example allows one to distinguish between the two ways of writing
this  sum rule, namely, keeping and omitting the elastic contribution.

The first way (which is the only meaningful one in long-range theories like
QED,
because both elastic and inelastic terms are IR divergent) leads to the
smooth interpolation between low and high $Q^2$.
This observation supports the suggestion of X. Ji \cite{Ji} to consider
such a quantity for the interpolation between high and low $Q^2$
for the real proton.
However, such an approach has nothing to do
with the original GDH value for real photons, which is
changed comletely by the infinitely growing elastic background.

At the same time, in a short-range theory with a mass gap (finite
threshold),
like QED with IR cutoff (or real QCD, where it is implied
by the confinement property), the interpolation between inelastic
contribution at non-zero $Q^2$ and the GDH value at $Q^2=0$ is possible. 
The form factor $I_{1+2}$ is rather smooth.
Our simple model is thus supporting
the hypothesis about the dominant role of the $g_2(x,Q^2)$ structure function
in the $Q^2$ dependence of the GDH sum rule.
However, further investigations
in the framework of, say, chiral models and/or QCD sum rules are highly
desirable. It would be especially interesting to obtain a direct
quantitative
estimate for the relevant cat-ear contributions, e.g. from
lattice gauge calculations.\\

We are indebted to B.L. Ioffe and A.V. Radyushkin
for useful discussions and valuable comments.
I.M. was supported by the
US Department of Energy  under contract DE-AC05-84ER40150
and grant DE-FG05-94ER40832. O.T. was supported by the Russian
Foundation for the Fundamental Research (grant 96-02-17631)
and the Graduiertenkolleg Erlangen-Regensburg (DFG).

\bb{99}
\bi{Ger} S.B.Gerasimov, Yad. Fiz. {\bf 2}, 598(1965)
[Sov. J. Nucl Phys. {\bf 2}, 430(1966)].
\bi{DH} S.D.Drell and A.C.Hearn, Phys. Rev. Lett. {\bf 16}, 908(1966).
\bi{E143} E143Collaboration, K.Abe et al.,Phys.Rev.Lett.{\bf78},815
(1997).
\bi{SoTe} J.Soffer and O.Teryaev, Phys. Rev. Lett. {\bf 70}, 3373(1993).
\bi{ST95} J.Soffer and O.Teryaev, Phys. Rev. {\bf 51}, 25(1995).
\bi{Ioffe} V.D.Burkert and B.L.Ioffe, Phys.Lett.{\bf B296}, 223(1992);
\bi{IoLi} B.L.Ioffe, V.A.Khoze, L.N.Lipatov, {\it Hard Processes},
North-Holland, 1984.
\bi{Sch} J.Schwinger, Proc. Natl. Acad. Sci. U.S.A. {\bf 72}, 1559(1975).
\bi{Mil} Wu-Yang Tsai, L.DeRaad and K.A.Milton, Phys. Rev. {\bf D11},
3537(1975).
\bi{AIL} M. Anselmino, B.L. Ioffe and E. Leader, Yad. Fiz. {\bf 49}
(1989) 214.
\bi{DYW} S.D. Drell and T.M. Yan, Phys. Rev. Lett. 24 (1970)181\\
G.B. West Phys. Rev. Lett. 24 (1970) 1206
\bi{BC} H.Burkhardt and W.N.Cottingham, Ann. Phys. (N.Y.) {\bf 16}, 543(1970).
\bi{Mer-vN} R.Mertig, W.L. van Neerven, Z. Phys. {\bf C60}, 489(1993).
\bi{Alt} G. Altarelli et al., Phys.Lett.{\bf B334}, 187 (1994).
\bi{Ji} Xiangdong Ji, Phys. Lett. {\bf B309}, 187 (1993).
\eb

\newpage
\begin{center}
{\large \bf Figure caption}
\end{center}

\noindent
{\bf Fig.1} \\[0.5cm]
\noindent

$4 I_{1+2}/\mu_A$ as a function of $y= Q^2/m^2$ for different values of the
threshold $\lambda$: a) $\lambda = 0.1$ (solid), b) $\lambda = 0.3$
(dashed), c) $\lambda = 1.0$ (dotted); $e=1$.

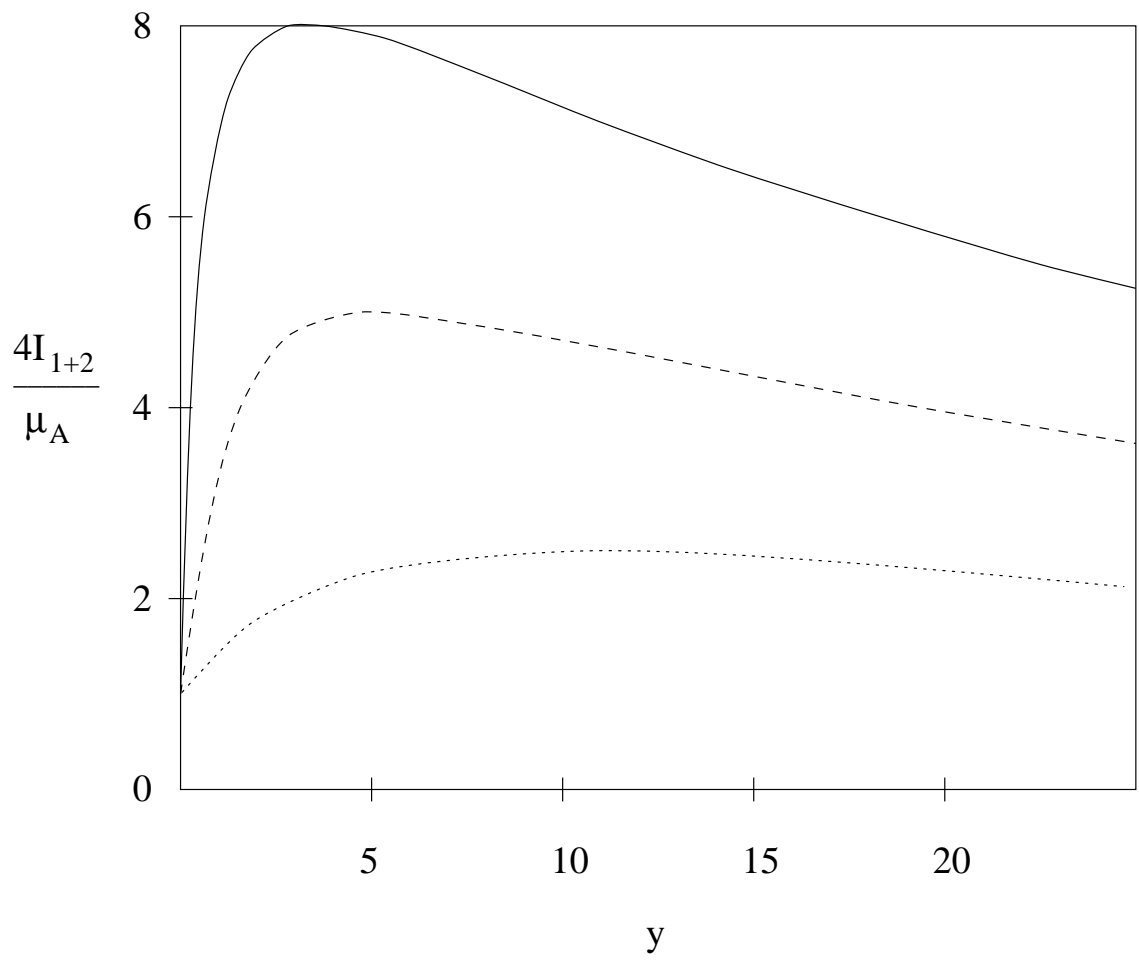
\end{document}